\documentclass[a4paper,14pt]{article}
\usepackage[english]{babel}
\usepackage{amsmath}
\usepackage{ulem}
\usepackage{calc}
\usepackage{vmargin}
\usepackage{amstext}
\usepackage{graphicx}
\usepackage{amsfonts}
\usepackage{amssymb}

\begin{document}
	
	\large	
	
	\title{
		{\bf Solutions of Three-Dimensional Stationary Gas Dynamics Equations} }
	

	\author{ \bf O.V. Kaptsov
		\\ Federal Research Center for Information and Computing Technologies,\\
		Novosibirsk, Russia\\
		E-mail: profkap@mail.ru}
	
	\date{}
	\maketitle
	
	This paper examines the three-dimensional stationary equations of a polytropic gas and employs symmetry methods to construct exact analytical solutions. In the Chaplygin gas case, the analysis yields a highly general solution family depending on three arbitrary functions, while the general adiabatic index formulation admits explicit solutions parameterized by several constants.

	
	\numberwithin{equation}{section}
	
	\section{Introduction} 
	
	The role of symmetries in physics is well known; they have proven to be extremely useful in quantum mechanics, special relativity, and field theory.
	Comprehensive treatments of symmetry methods for nonlinear partial differential equations can be found in the monographs \cite{OvsGA, Ibragimov, Olver}. 
	Applications of symmetries to hydrodynamic equations are discussed extensively in the recent monographs \cite{AndreevKaptsov, Badin, Pukhnachev}.
	In 1994, L.V.~Ovsyannikov \cite{OvsPodmod} introduced the “Submodel Program,” aimed at studying solutions of reduced hydrodynamic models.
	A central goal of this program is the construction of solutions to the equations of gas dynamics.
	Despite significant efforts, explicit solutions of nonlinear three-dimensional hydrodynamic equations remain scarce.
	
	This work is devoted to constructing solutions of the three-dimensional stationary equations of gas dynamics.
	Section~2 considers potential three-dimensional flows of an ideal gas.
	We show that if the entropy and the Bernoulli integral are constant and the flow is sonic, then the corresponding state necessarily satisfies a polytropic law with index $-1$, which corresponds to the Chaplygin gas.
	We then derive the symmetry algebra associated with the equation
	\[
	\operatorname{div}\!\left(\frac{\nabla \phi}{|\nabla \phi|}\right)=0,
	\]
	which describes the Chaplygin gas in terms of the velocity potential~$\phi$.
	This algebra is infinite-dimensional; the coefficients of its generators depend on eight arbitrary functions.
	Such a structure enables the construction of solutions involving three arbitrary functions.
	
	Section~3 examines stationary vortical flows.
	Following the approach of Refs.~\cite{AndreevKaptsov,KaptsovArt} and applying an additional dilatation transformation, we reduce the governing equations in both the axisymmetric and fully three-dimensional settings to a system of ordinary differential equations.
	The reduced system possesses a sufficient number of first integrals to further reduce it to a single differential equation.
	In a particular case, explicit solutions of this equation are obtained.
	This leads to closed-form stationary solutions of the gas dynamic equations expressed in elementary functions, together with the corresponding streamline formulas.

\section{Potential Flows and Chaplygin Gas.}   

We begin with the three-dimensional steady Euler equations for an ideal gas.
In the absence of external forces, the governing equations comprise the
continuity equation, the momentum balance, and the energy relation.  
In compact vector form, these equations can be written as
\[
\begin{aligned}
	\nabla\!\cdot(\rho \mathbf{u}) &= 0, \\[1.8mm]
	\rho\,(\mathbf{u}\!\cdot\nabla)\mathbf{u} + \nabla p &= \mathbf{0}, \\[1.8mm]
	\mathbf{u}\!\cdot\nabla p + c^{2}\rho\,\nabla\!\cdot\mathbf{u} &= 0 ,
\end{aligned}
\]
where $\rho$ is the density, $\mathbf{u}=(u,v,w)$ is the velocity field,
$p$ is the pressure, and $c$ is the local sound speed.

Let \(q = u^{2} + v^{2} + w^{2}\)
denote the square of the flow speed.
We assume that the Bernoulli integral
\begin{equation}
	I_B = \frac{q}{2} + \int \frac{c^{2}, d\rho}{\rho}
\end{equation}
and the entropy remain constant throughout the flow domain.
In addition, we consider the case in which the flow speed equals the local sound speed,
\begin{equation}
	q = c^{2}.
\end{equation}
These assumptions imply the relation
\begin{equation} \label{I2}
	\frac{q}{2} + \int \frac{q}{\rho}\, d\rho = k, 
	\qquad k \in \mathbb{R}.
\end{equation}
Because \(q\) depends solely on \(\rho\), differentiation of \eqref{I2} with respect to \(\rho\)
yields the ordinary differential equation
\[
q' + \frac{2q}{\rho} = 0 .
\]
Its solution is
\begin{equation} \label{q}
	q = \frac{a}{\rho^{2}},
\end{equation}
where \(a > 0\) is an arbitrary constant.
Substitute this expression into the Bernoulli integral
\[
\frac{a}{2\rho^{2}} + \int \frac{p'\, d\rho}{\rho} = k,
\]
and differentiate the resulting relation with respect to \(\rho\). This gives the differential equation
\[
\frac{p'}{\rho} = \frac{a}{\rho^{3}}.
\]
Its solution,
\begin{equation} \label{p=}
	p = -\frac{a}{\rho} + b, 
	\qquad b \in \mathbb{R},
\end{equation}
defines the equation of state of the Chaplygin gas \cite{Mises}.

Thus, the assumptions that the flow speed equals the sound speed and that both the Bernoulli integral and the entropy remain constant imply that the gas must obey the Chaplygin equation of state.

We now turn to potential flows, i.e., we assume the existence of a scalar function \(\phi\) such that its gradient equals the velocity vector \(V = (u, v, w)\). As is well known \cite{Mises}, the potential then satisfies
\[
(\rho \phi_x)_x + (\rho \phi_y)_y + (\rho \phi_z)_z = 0.
\]
For the Chaplygin gas, this equation reduces to
\begin{equation} \label{div}
	\operatorname{div}\!\left(\frac{\nabla \phi}{|\nabla \phi|}\right) = 0.
\end{equation}
In explicit form, equation \eqref{div} becomes
\begin{equation} \label{Pot}
	(\phi_y^{2} + \phi_z^{2})\, \phi_{xx}
	+ (\phi_x^{2} + \phi_z^{2})\, \phi_{yy}
	+ (\phi_x^{2} + \phi_y^{2})\, \phi_{zz}
	- 2(\phi_x \phi_y \phi_{xy}
	+ \phi_x \phi_z \phi_{xz}
	+ \phi_y \phi_z \phi_{yz}) = 0.
\end{equation}

It is straightforward to verify that the corresponding equation for characteristics can be represented as
\begin{equation} \label{har}
	(\phi_x \Phi_y - \phi_y \Phi_x)^2 + (\phi_x \Phi_z - \phi_z \Phi_x)^2 + (\phi_y \Phi_z - \phi_z \Phi_y)^2 = 0.
\end{equation}
The last equation for the function $\Phi$ must hold on solutions of equation \eqref{Pot}. Obviously, if $\phi$ is a solution of equation \eqref{Pot}, then $\Phi = \phi$ satisfies \eqref{har}. Hence, the relation
$\phi = \text{const}$
defines the characteristics.

Let us proceed to constructing solutions to equation \eqref{Pot} for the potential. First, we find the symmetries of equation \eqref{Pot}. Applying standard methods of group analysis
\cite{OvsGA,Olver}, we find the infinite-dimensional Lie algebra of admissible operators
$$  X = (F_7 x - F_5 z - y F_2 + F_4) \frac{\partial}{\partial x} + (F_7 y - F_6 z + x F_2 + F_3) \frac{\partial}{\partial y} + $$
$$ (F_5 x + F_6 y + z F_7 + F_8) \frac{\partial}{\partial z} + F_1 \frac{\partial}{\partial \phi},  $$ 
where $F_1, \dots, F_8$ are arbitrary functions of $\phi$.

Take two operators
$$ X_1 = F_4 \frac{\partial}{\partial x} + F_3 \frac{\partial}{\partial y}, \qquad
X_2 = F_9 \frac{\partial}{\partial y} + F_8 \frac{\partial}{\partial z}, $$
where $F_9$ is an arbitrary function of $\phi$.
For finding invariants, it is necessary to solve the system of first-order linear differential equations \cite{OvsGA}
\begin{equation} \label{Inv}
	X_1 I = 0, \qquad X_2 I = 0 .
\end{equation}	
Obviously, $I_1 = \phi$ satisfies the system \eqref{Inv}. The function $I_2 = a x + b y + c z$ is a solution if the functions $a, b, c$ satisfy the system of algebraic equations 
$$ F_4 a + F_3 b = 0, \qquad F_9 b + F_8 c = 0. $$
Therefore, any smooth function $G$ of $I_1$, $I_2$ is an invariant. Equating $G$ to zero, we get a representation for solutions of equation \eqref{Pot}. 
By direct calculation, it is not difficult to verify that the formula
\begin{equation} \label{solution}
	a x + b y + c z + d = 0,  
\end{equation}      
where $a, b, c, d$ are arbitrary functions of $\phi$,
indeed defines an implicit solution of the potential equation \eqref{Pot}.
Note that the actual arbitrariness in the solution is three functions.

From representation \eqref{solution}, one can obtain various explicit solutions.
As an example, consider the potential
\begin{equation} \label{Ssol}
	\phi = f!\left(\frac{k_1 x + k_2 y + k_3 z + k_4}{n_1 x + n_2 y + n_3 z + n_4}\right),
\end{equation}
where $k_i, n_j$ are arbitrary constants, and $f$ is any smooth function.
 If $n_1 = \cdots = n_3 = 0$, $n_4 = 1$, then the potential is smooth.
 
 Obviously, the obtained solutions \eqref{solution} correspond to characteristic surfaces which are planes. In the two-dimensional case, characteristics are straight lines. 
In general, the intersection of characteristic planes (characteristic lines) leads to the formation of shock waves. If the planes intersect along a line, we get solutions of the centered wave type. If the planes do not intersect, solutions of equation \eqref{Pot} can be smooth.

\section{Vortical Flows.} 
In this section, we first examine stationary adiabatic axisymmetric flows of an ideal gas, which are governed by the system of equations  
\begin{equation} \label{Sgd}
	\begin{cases}
		u\rho_z + v\rho_r + \rho\!\left(u_z + v_r + \dfrac{v}{r}\right) = 0, \\[4pt]
		\rho (u u_z + v u_r) + p_z = 0, \\[4pt]
		\rho (u v_z + v v_r) + p_r = 0, \\[4pt]
		u p_z + v p_r + \gamma p\!\left(u_z + v_r + \dfrac{v}{r}\right) = 0,
	\end{cases}
\end{equation}
where $u$ and $v$ are the velocity components, $\rho$ is the density, and $p$ is the pressure.

It can be shown that the Lie algebra of point symmetries of system~\eqref{Sgd}
is generated by the operators
\begin{equation*}
	X_1 = \frac{\partial}{\partial z}, \qquad
	X_2 = p\,\frac{\partial}{\partial p} + \rho\,\frac{\partial}{\partial \rho}, \qquad
	X_3 = z\,\frac{\partial}{\partial z} + r\,\frac{\partial}{\partial r},
\end{equation*}
together with an infinite family of operators of the form
\begin{equation} \label{Y}
	Y_{\mathcal{F}} = \mathcal{F}(I_1, I_2)\!\left(
	u\,\frac{\partial}{\partial u}
	+ v\,\frac{\partial}{\partial v}
	- 2\rho\,\frac{\partial}{\partial \rho}
	\right),
\end{equation}
Here $\mathcal{F}$ is an arbitrary function of entropy and the Bernoulli integral
$$ I_1 = \frac{p}{\rho^\gamma}, \qquad I_2 = \frac{u^2 + v^2}{2} + \frac{\gamma p}{(\gamma - 1) \rho}. $$
An operator of the form \eqref{Y} in the two-dimensional and three-dimensional cases is presented in \cite{OvsGasDyn, Meleshko}.

Now we assume that the pressure does not depend on $z$. Then the system \eqref{Sgd} admits an additional dilation operator
$$ X_5 = u\frac{\partial}{\partial u} + z\frac{\partial}{\partial z} . $$
Consider a linear combination of operators $Z = X_5 + m Y_{\mathcal{F}}$, where $m \in \mathbb{R}$, $\mathcal{F} = 1$. The invariant solution with respect to the operator $Z$
must have the form
\begin{equation} \label{IS}
	u = z^{m+1} U, \quad v = z^m V, \quad \rho = z^{-2m} Q, \quad p = P,
\end{equation}	
where $U, V, Q, P$ are functions of $r$.
Substituting the representation \eqref{IS} into \eqref{Sgd}, we obtain a system of ordinary differential equations
\begin{align}
&(1-m) U Q + V Q^{\prime} + Q V^{\prime} + \frac{V Q}{r} = 0,\label{33}\\
&	(1+m) U^2 + V U^{\prime} = 0,\label{34}\\	
&	m V U + V V^{\prime} + \frac{P^{\prime}}{Q} = 0, \label{35}\\
&		V P^{\prime} + \gamma P \left( (m+1) U + V^{\prime} + \frac{V}{r} \right) = 0.\label{36}
\end{align}


Rewrite equation \eqref{34} in the form
\begin{equation} \label{UV}
	\frac{U}{V} = -\frac{U^{\prime}}{(m+1) U}.
\end{equation}
Divide equation~\eqref{33} by $Q V$ and substitute expression~\eqref{UV} into the resulting relation.
As a result, equation~\eqref{33} can be rewritten as
$$ \frac{m-1}{m+1} (\ln U)^{\prime} + (\ln Q V)^{\prime} + (\ln r)^{\prime} = 0. $$
From here we find the first integral
$$ Q V r U^{\frac{m-1}{m+1}} = c_1, \quad c_1 \in \mathbb{R}. $$
Thus, we obtain the representation
\begin{equation} \label{QUV}
	Q = \frac{c_1 U^{\frac{1-m}{m+1}}}{V r}.
\end{equation}
Similarly, if we divide equation \eqref{36} by $P V$ and use formula \eqref{UV}, then \eqref{36} can be written as
$$ \left( \ln \big( P V^{\gamma} r^{\gamma} U^{-\gamma} \big) \right)^{\prime} = 0. $$
From here we obtain the first integral and the representation for the function $P$
\begin{equation} \label{PUV}
	P = c_2 \left( \frac{U}{V r} \right)^{\gamma},
\end{equation}
where $c_2$ is an arbitrary constant.

Next, we divide equation \eqref{35} by $V^2$ and rewrite it as
$$ -\frac{m}{m+1} \frac{U^{\prime}}{U} + \frac{V^{\prime}}{V} + \frac{P^{\prime} r V}{Q V^2} = 0. $$
Substituting the representation \eqref{QUV} into the last expression, we obtain the equation
$$ \frac{(V U^{-\frac{m}{m+1}})^{\prime}}{V U^{-\frac{m}{m+1}}} + \frac{P^{\prime} r V}{c_1 U^{\frac{1-m}{m+1}} V^2} = 0. $$
This equation can be written as
$$ (V U^{-\frac{m}{m+1}})^{\prime} + \frac{P^{\prime} r V U^{-1}}{c_1 V U^{-\frac{m}{m+1}}} = 0. $$
Using the representation \eqref{PUV}, we find the first integral
\begin{equation} \label{FI2}
	\frac{V^2 U^{\frac{-2m}{m+1}}}{2} + \frac{c_2 \gamma}{c_1 (\gamma - 1)} \left( \frac{U}{V r} \right)^{\gamma - 1} = c_3,
	\qquad c_3 \in \mathbb{R}.
\end{equation}
Since
\begin{equation} \label{V=UI}
	V = -(m+1) \frac{U^2}{U^{\prime}},
\end{equation}
the first integral \eqref{FI2} gives an ordinary differential equation for the function $U$
\begin{equation} \label{DU}
	\frac{(1+m)^2}{2} \frac{U^{\frac{2m+4}{m+1}}}{(U^{\prime})^2} + \frac{c_2 \gamma}{c_1 (\gamma - 1)} \left( \frac{-U^{\prime}}{(m+1) U r} \right)^{\gamma - 1} = c_3.
\end{equation}
To find the streamlines, one needs to integrate the equation
\begin{equation} \label{CL}
	\frac{dz}{u} = \frac{dr}{v}.
\end{equation}
Using the representations \eqref{IS} and \eqref{V=UI}, the equation for the streamlines can be rewritten as
$$ \frac{dz}{z} = \frac{U^{\prime} dr}{-(m+1) U}. $$
Hence, the streamlines are given by the formula
$$ z = c U^{-\frac{1}{m+1}}, $$
where $c$ is an arbitrary constant.

It remains to find the solution of equation~\eqref{DU}. However,
for an arbitrary value of $\gamma$, equation~\eqref{DU} cannot be solved explicitly when $c_3 \neq 0$.
Assume that $c_3 = 0$. Then, according to~\eqref{DU}, the function $U$ takes the form
\[
U = \left( a\, r^{\frac{2\gamma}{1+\gamma}} + b \right)^{-\frac{(m+1)(1+\gamma)}{2}} .
\]
Here, $b$ is an arbitrary constant, and $a$ is determined by the constants appearing in equation~\eqref{DU}.

Following~\cite{KaptsovArt}, let us now assume that, in system~\eqref{Sgd}, the derivative $p_r$ is equal to zero.
In this case, system~\eqref{Sgd} admits an additional scaling operator  
\[
Z = r\,\frac{\partial}{\partial r} + v\,\frac{\partial}{\partial v} .
\]
Using a linear combination of the operators $Y$ and $Z$, it is easy to obtain the following representation for the solutions \cite{KaptsovArt}
\begin{equation} \label{sol1} 
	u = r^m U , \quad v = r^{m+1} V , \quad \rho = r^{-2m} Q , \quad p = P ,     
\end{equation}
where $U, V, Q, P$ are some functions of $z$, and $m \in \mathbb{R}$. 
Substituting this representation into the system \eqref{Sgd}, we obtain four ordinary differential equations of the type \eqref{33}--\eqref{36}.  
The system of four equations admits two first integrals \cite{KaptsovArt}:
$$ Q U V^{\frac{2-m}{1+m}} = c_1, \quad 
\left( \frac{U}{V^{\frac{m+2}{m+1}}} \right)^{\gamma} P = c_2 , $$
where $c_1, c_2$ are arbitrary constants.  
Using these integrals, one can obtain a third integral -- a differential equation for the function $V$
\begin{equation} \label{FI}   
	\frac{1}{2(m+1)^2} \left(\frac{V^{\frac{m+2}{m+1}}}{V^{\prime}} \right)^2 +
	\frac{c_2\gamma}{c_1(\gamma-1)} \left(-(1+m)V^{\frac{-m}{m+1}} V^{\prime}  \right)^{\gamma-1} = c_3 , 
	\qquad c_3 \in \mathbb{R} . 
\end{equation}
To derive this equation, the following relation was used
$$ U = \frac{-V^2}{(m+1)V^{\prime}} \ . $$
Integrating the differential equation \eqref{CL}, we obtain a representation for the streamlines 
in terms of the function $V$:
$$  r = c V^{\frac{-1}{m+1}}  \ , \quad c \in \mathbb{R} . $$

Let us assume that $c_3 = 0$ in equation \eqref{FI}. 
Then this equation has two types of solutions.  
For $\gamma = 3$, the solution takes the form
\begin{equation} \label{V1}
	V = b e^{kz} , \qquad k = \frac{-1}{(m+1)} \left( \frac{c_1}{3c_2} \right)^{\frac{1}{4}} , \qquad b \in \mathbb{R} ,   
\end{equation}
and for $\gamma \neq 3$,
$$ V = (a z + b)^{\frac{\gamma - 3}{(\gamma + 1)(m + 1)}} \ , $$
where $b$ is an arbitrary constant, and $a$ is expressed in terms of the constants in equation \eqref{FI}.  
The solution \eqref{sol1} corresponding to the function \eqref{V1} is smooth.  
If one of the streamlines is considered as a solid boundary, then this solution can be interpreted as the flow around a curved wall.

Let us now consider the three-dimensional stationary gas dynamics equations  

\begin{align}
&u\rho_x + v\rho_y + w\rho_z + \rho \left(u_x + v_y + w_z \right) = 0,  \notag\\
&\rho (u u_x + v u_y + w u_z) + p_x = 0, \notag \\
&\rho (u v_x + v v_y + w v_z) + p_y = 0,  \label{GD3d} \\
&\rho (u w_x + v w_y + w w_z) + p_z = 0,  \notag\\
&u p_x + v p_y + w p_z + \gamma p (u_x + v_y + w_z) = 0 . \notag
\end{align}

%
By analogy with the axisymmetric case, it is easy to verify \cite{OvsGasDyn} that the system \eqref{GD3d} admits a symmetry operator of the form  
$$
Y_{ \mathcal{F}} = \mathcal{F}(I_1, I_2) \cdot \left( 
u\frac{\partial}{\partial u} + v\frac{\partial}{\partial v }  -
2\rho\frac{\partial}{\partial \rho }
\right) ,
$$
where $\mathcal{F}$ is an arbitrary function of the entropy and the Bernoulli integral,
$$ 
I_1 = \frac{p}{\rho^{\gamma}} , \qquad 
I_2 = \frac{u^2 + v^2 + w^2}{2} + \frac{\gamma p}{(\gamma - 1)\rho} .
$$

If we assume that the pressure $p$ does not depend on $x$, then the system \eqref{GD3d} admits
a scaling operator
\begin{equation} \label{Y3}
	x\frac{\partial}{\partial x }
	+ (m+1)u\frac{\partial}{\partial u}
	+ m v\frac{\partial}{\partial v }
	+ m w\frac{\partial}{\partial w }
	- 2m\rho\frac{\partial}{\partial \rho } , 
\end{equation}
where $m$ is an arbitrary constant.  
This operator allows us to obtain the following representation for the solutions:
$$
u = x^{m+1}\mathcal{U}, \quad
v = x^{m}\mathcal{V}, \quad
w = x^{m}\mathcal{W}, \quad
\rho = x^{-2m}\mathcal{Q}, \quad
p = \mathcal{P},
$$   
where $\mathcal{U}, \mathcal{V}, \mathcal{W}, \mathcal{Q}, \mathcal{P}$ are some functions of $y$ and $z$.  
If we substitute this representation into the system \eqref{GD3d}, we obtain a system of five equations for these functions.  
The resulting system also admits a scaling transformation, which makes it possible to seek a solution of \eqref{GD3d} in the form
\begin{equation} \label{sol} 
	u = x^{m+1}y^nU ,\quad 
	v = x^{m}y^{n+1}V ,\quad 
	w = x^{m}y^nW ,\quad 
	\rho = x^{-2m}y^{-2n}Q , \quad 
	p = P ,     
\end{equation}
where $U, V, W, Q, P$ are functions of $z$.  
Substituting the representation \eqref{sol} into \eqref{GD3d}, we obtain
a system of five ordinary differential equations
\begin{align}
(1 - n) Q U + (1 - m) Q V + (QW)^{\prime} = 0 ,\label{E1}\\
	(1 + n) U^2 + m U V + W U^{\prime} = 0 ,\label{E2}\\
	n U V + (1 + m) V^2 + W V^{\prime} = 0 ,\label{E3}\\
	Q (n U W + m V W + W W^{\prime}) + P^{\prime} = 0 ,\label{E4}\\
		W P^{\prime} + \gamma ((n + 1) U + (m + 1) V + W^{\prime}) = 0 .\label{E5}
\end{align}

Let us introduce the notation
$$ S = U^{\frac{1}{1+ n + m}} , \qquad R = V^{\frac{1}{1 + n + m}} . $$
Then, from equations \eqref{E2} and \eqref{E3}, we obtain
\begin{equation} \label{UVW}
	\frac{V}{W} = \left( \frac{S^n}{R^{n+1}} \right)^{\prime} , \qquad
	\frac{U}{W} = \left( \frac{R^m}{S^{m+1}} \right)^{\prime} .
\end{equation}
As shown in \cite{AndreevKaptsov}, the system \eqref{E1}--\eqref{E5} admits three first integrals
\begin{equation} \label{c1c2}
	Q W R^{2m - 1 - n} S^{2n - 1 - m} = c_1 , \qquad 
	P \left( W S^{-1 - m} R^{-1 - n} \right)^{\gamma} = c_2 ,
\end{equation}
\begin{equation} \label{fint} 
	W^2 R^{-2m} S^{-2n} + \frac{2 c_2 \gamma}{c_1 (\gamma - 1)} 
	\left( \frac{R^{n+1} S^{m+1}}{W} \right)^{\gamma - 1} = c_3 ,
\end{equation}
where $c_1$, $c_2$, and $c_3$ are arbitrary constants.  

Following \cite{AndreevKaptsov}, let us introduce additional notation
\begin{equation}\label{X,Y}
	T = \frac{R^{n+1} S^{m+1}}{W} , \quad 
	X = S^n R^{-n-1} , \quad 
	Y = R^m S^{-m-1} . 
\end{equation}
Then equations \eqref{UVW} can be rewritten as
$$ \frac{U}{W} = \frac{Y^{\prime}}{Y} , \qquad \frac{V}{W} = \frac{X^{\prime}}{X} . $$
From the last equations, it is easy to obtain two new relations
$$ Y^{\prime} = X Y T , \qquad X^{\prime} = X Y T . $$
Hence,
$$ Y = X + c_4 , \quad T = \frac{X^{\prime}}{X(X + c_4)} , \qquad c_4 \in \mathbb{R} . $$
The first integral \eqref{fint} in the new variables takes the form  
$$ \frac{1}{2 X^2 (X + c_4)^2 T^2} + a T^{\gamma - 1} = c_3 , $$
where $a$ is expressed in terms of $c_1$, $c_2$, and $\gamma$.  
From the last two expressions, we obtain an ordinary differential equation for the function $X$
\begin{equation} \label{Xprime}
	\frac{1}{(2 X^{\prime})^2} + a \left( \frac{X^{\prime}}{X (X + c_4)} \right)^{\gamma - 1} = c_3 .
\end{equation}

For an arbitrary constant $c_3$, it is not possible to find an explicit solution of the equation \eqref{Xprime}. 
Therefore, we assume that $c_3 = 0$.  
If $\gamma = 3$, then the differential equation \eqref{Xprime} takes the form
$$ X^{\prime} + A \sqrt{X (X + c_4)} = 0 , $$
where $A = (2a)^{\frac{-1}{\gamma + 1}}$.  
The solution to this  equation is
$$ X = \frac{(2 e^{-A z + b} - c_4)^2}{8 e^{-A z + b}} , \qquad b \in \mathbb{R} . $$

Now let $c_3 = c_4 = 0$ and $\gamma \neq 3$. Then equation \eqref{Xprime} can be written as
$$ X^{\prime} + A X^{\frac{2(\gamma - 1)}{\gamma + 1}} = 0 . $$
Its solution is
$$ X = \left( \frac{A (\gamma - 3)(z + b)}{\gamma + 1} \right)^{\frac{-\gamma - 1}{\gamma - 3}} , 
\qquad \forall b \in \mathbb{R}. $$
Since
$$ Y = X , \qquad T = \frac{X^{\prime}}{X^2} , $$
then, from relations \eqref{X,Y}, we find
$$ 
S = R = \frac{1}{X} = 
\left( \frac{A (\gamma - 3)(z + b)}{\gamma + 1} \right)^{\frac{\gamma + 1}{\gamma - 3}} , \qquad
T = A \left( \frac{A (\gamma - 3)(z + b)}{\gamma + 1} \right)^{\frac{4}{\gamma - 3}} .
$$
Hence, the functions $U$ and $V$ are equal:
$$ 
U = V = S^{1 + m + n} =
\left( \frac{A (\gamma - 3)(z + b)}{\gamma + 1} \right)^{\frac{(\gamma + 1)(1 + n + m)}{\gamma - 3}} .
$$
According to the first formula in \eqref{X,Y}, we obtain
$$ 
W = \frac{R^{n + 1} S^{m + 1}}{T} =
\frac{z (\gamma - 3)}{\gamma + 1}
\left( \frac{A (\gamma - 3)(z + b)}{\gamma + 1} \right)^{\frac{(\gamma + 1)(1 + n + m)}{\gamma - 3}} .
$$
The first integrals \eqref{c1c2} allow us to find
$$ 
P = c_2 T^{\gamma} , \qquad 
Q = c_1 T R^{-2m} S^{-2n} .
$$
Since the streamlines are defined by the equations
$$ 
\frac{dx}{u} = \frac{dy}{v} = \frac{dz}{w} ,
$$ 
in this case these equations reduce to   
$$ 
\frac{dx}{x} = \frac{dy}{y} = \frac{(\gamma + 1) dz}{(\gamma - 3) z} .
$$
The parametric representation of the streamlines is 
$$ 
x = a_1 e^t , \quad 
y = a_2 e^t , \quad 
z = a_3 e^{t \frac{(\gamma - 3)}{\gamma + 1}} , 
\qquad a_1, a_2, a_3 \in \mathbb{R} .
$$

\section{Conclusion}

In this study, new classes of solutions to the three-dimensional stationary gas dynamics equations have been obtained.  
Such solutions are exceptionally rare in the existing literature.  
For the Chaplygin gas, we derived a family of solutions depending on three arbitrary functions.  
It can also be shown that, in the case of an arbitrary spatial dimension $n$, the nonlinear equation \eqref{div} possesses an infinite-dimensional symmetry algebra.  
This property makes it possible to construct implicit solutions of the form
\[
a_1 x_1 + \dots + a_n x_n + b = 0,
\]
where $a_1, \dots, a_n,$ and $b$ are arbitrary functions of the potential $\phi$.  
Such solutions can be obtained more directly using the $B$-determining equation method \cite{AndreevKaptsov}; however, in the present work, classical Lie symmetry techniques were employed for methodological clarity and for comparison with previously established approaches.

The vortex-type solutions presented in Section~3 were obtained using an additional stretching symmetry operator.  
Although first integrals of the reduced systems of ordinary differential equations had been derived in earlier studies \cite{AndreevKaptsov,KaptsovArt}, explicit closed-form analytical expressions for the corresponding solutions were not available.  
The results reported here therefore extend those works by providing explicit functional forms and associated streamline equations.  
It is also worth noting that the existence of additional symmetry operators was discussed in the monograph \cite{AndreevKaptsov}, where they were obtained via the $B$-determining equation framework.  
This approach can be generalized to other classes of nonlinear hydrodynamic equations \cite{KaptsovBook}, making it a promising analytical tool for constructing exact solutions of new mathematical models of fluid dynamics.

\bigskip
\noindent\textbf{Acknowledgments.}  

The research was carried out within the state assignment of Ministry of Science and Higher Education of the Russian Federation for Federal Research Center for Information and Computational Technologies

This work is supported by the Krasnoyarsk Mathematical Center and financed by the Ministry of Science and Higher Education of the Russian Federation in the framework of the establishment and development of regional Centers for Mathematics Research and Education (Agreement No. 075-02-2025-1606).


\begin{thebibliography}{33}
	

	\bibitem{OvsGA}

Ovsiannikov L. V., Group Analysis of Differential Equations. Academic Press. NY.   1982
	
	\bibitem{Ibragimov}
 Ibragimov N. H. Transformation Groups Applied to Mathematical Physics. Reidel Publishing Company. Boston. 1985
	
\bibitem{Olver}	
	Olver P. Applications of Lie groups to differential equations. Springer, NY. 1986
	
	\bibitem{AndreevKaptsov}
	Andreev V.K., Kaptsov O.V.,  Pukhnachev V. V.,  Rodionov A.A. 
	Applications of Group-Theoretical Methods in Hydrodynamics. Springer, NY. 2010
	
	
	\bibitem{Badin}	
	Badin G., Crisciani F. Variational Formulation of Fluid and Geophysical Fluid Dynamics Mechanics, Symmetries and Conservation Laws. Springer International Publishing.  2018
	
	\bibitem{Pukhnachev}
Pukhnachev V.V. Symmetries in the Navier-Stokes Equations. Novosibirsk: NSU, 2022
(in Russian)
	
	\bibitem{OvsPodmod}
Ovsyannikov L.V. The SUBMODELS program. Gas dynamics // J. Appl. Maihs Mechs, 1994. Vol. 58, No. 4, pp. 601-627

	
	\bibitem{KaptsovArt}
Kaptsov, O. V. Exact solutions of the equations of axisymmetric flow of an ideal gas // Dynamics of a continuous medium. Mathematical modeling. - Novosibirsk: IG SB RAS USSR, 1989. - Issue 91. - P. 37-47 (in Russian)
	
	\bibitem{Mises}
Mises, R. Mathematical Theory of Compressible Fluid Flow. Dover Publications, New York, 2004.
	
	\bibitem{OvsGasDyn}

	Ovsiannikov, L.V. Lectures on Basis of the Gas Dynamics, 2nd ed.; Institute of Computer Studies: Moscow-Izhevsk , Russia, 2003.
	
	\bibitem{Meleshko}
	Meleshko  S. V., Kaptsov E.I. Symmetry Analysis of the Two-Dimensional Stationary Gas
	Dynamics Equations in Lagrangian Coordinates// Mathematics, 2024, 12(6), 879
	
	\bibitem{KaptsovBook}
	Kaptsov, O. V. Methods of Integration of Partial Differential Equations. Moscow: FIZMATLIT, 2009. (in Russian)
	
\end{thebibliography}
\end{document}